\documentclass[12pt]{article}
\input epsf.sty

\def\lromn#1{\uppercase\expandafter{\romannumeral#1}}

\begin{document}

\begin{center}
\begin{large}
\textbf{
Macro-coherent two photon and radiative
neutrino pair emission
}

\end{large}
\end{center}

\vspace{2cm}
\begin{center}
\begin{large}
M. Yoshimura$^{\dagger}$, C. Ohae$^{\diamond}$,
A. Fukumi$^{\dagger}$, K. Nakajima$^{\dagger}$, \\
I. Nakano$^{\dagger}$$^{\diamond}$,
H. Nanjo$^{\ddagger}$,  and N. Sasao$^{\ddagger}$, 

$^{\dagger}$Center of Quantum Universe and
$^{\diamond}$Department of Physics, Okayama University, \\
Tsushima-naka 3-1-1, Okayama,
700-8530 Japan
\\
$^{\ddagger}$Department of Physics, Kyoto University,\\
Kitashirakawa, Sakyo, Kyoto,
606-8502 Japan 

\end{large}
\end{center}

\vspace{2cm}

\begin{center}
\begin{Large}
{\bf ABSTRACT}

\end{Large}
\end{center}

We discuss a possibility of detecting
a coherent photon pair
emission and related radiative neutrino pair emission
from excited atoms.
It is shown that atoms of lambda- and ladder-type
three level system placed in a pencil-like cylinder 
give a back to back emission
of two photons of equal energy $\Delta/2$, 
sharply peaked with a width $\propto $ 1/( target size) and
well collimated
along the cylinder axis.
This process has a measurable rate $\propto$ (target number density)
$^2 \times$ target volume, while a broader 
spectral feature of one-photon distribution 
separated by (mass sum of a neutrino pair)$^2/(2\Delta)$
from the two photon peak may arise from radiative neutrino pair emission, 
with a much smaller rate.

\vspace{2cm}

{\bf Introduction}

Superradiance proposed by Dicke 
\cite{dicke},\cite{super-review} is a remarkable effect,
giving a large rate, at its maximum 
$\propto N^2$ with $N $ the number of target atoms, 
much larger than the spontaneous emission rate $\propto N$.
It may give rise to an effective enhancement of
weaker rates of forbidden transitions  such as M1, E2
and even weak interaction process of neutrino
pair emission,
thus giving a possibility of measuring these small rates.

In the present work we consider two photon
emission $|1 \rangle \rightarrow |3 \rangle + \gamma
+ \gamma$ and radiative neutrino pair emission
$|1 \rangle \rightarrow |3 \rangle + \gamma
+ \nu_i \nu_j$
from coherently excited targets of three level atomic system;
$| 1 \rangle \,, |3\rangle$
and some intermediate state $| 2\rangle$.
(We denote three neutrino
mass eigenstates by $\nu_i \,, i = 1\,, 2\,, 3$.)
The latter neutrino process is related to laser
irradiated neutrino pair emission discussed in \cite{my-06}
aiming at the neutrino mass spectroscopy.
It is shown below that these processes have 
rates $\propto n^2 V$ ($n$ the target number density and $V$
the target volume) and striking kinematical
features of angular correlation
and energy spectrum, hence may be detectable,
if the rate is large enough.

A very sharp single photon peak
of two photon pair emission
is located at the half of the energy difference
$\Delta/2\,, \; \Delta= E_1 - E_3$, and its
angular distribution
is well collimated, for a large aspect ratio, to the 
cylinder axis.
It can thus be used for a precise determination of $\Delta/2$.
This means that we do not need an independent
experiment for measurement of the value $\Delta$
from atoms in a complex envirornment.
Moreover, the photon pair is highly correlated
(back to back, and so on), and has a spin correlation with
atomic angular momentum involved.
Thus, these pairs are ideal entangled states.

On the other hand,
the photon energy 
distribution arising from radiative
neutrino pair emission has a threshold in the vicinity
of the two photon peak, ranging in the continuous energy region
$\hbar \omega \leq \Delta/2
- (m_i + m_j)^2/(2\Delta)$, with
$m_i$ three neutrino mass values.
Numerically, $(2m_3)^2/2\Delta = 5 meV\,(m_3/50 meV)^2(0.1 eV/\Delta)$
for the heaviest neutrino $\nu_3$.
The threshold rise of the rate thus 
provides a critical information of neutrino masses, giving
their sum  $m_i + m_j$, if the process has a measurable rate.
It might even be possible in a distant future
to detect relic neutrino of 1.9 K \cite{my-taka},
if the neutrino mass spectroscopy works ideally.

As usual, we use the natural unit such that
$\hbar =1$ and $c=1$ throughout the present paper.
We abbreviate the new phenomenon of
coherent two photon pair emission
as macro-coherent two photon emission (MCTPE).

\vspace{1cm}
{\bf Macro-coherent two photon emission}

We consider a coherent collection of
excited atoms having three level structure
of lambda($\Lambda$)- and ladder-type.
An example of Ba atom levels is depicted in Figure 1.
In this example both of the two types  coexist;
$\Lambda-$type, 
$^1D_2(|1\rangle) - ^1P_1(|2\rangle) - ^1S_0(|3\rangle)$, 
and the ladder-type,
$^1D_2(|1\rangle) - ^3D_{2\,, 1\,, 0}(|2\rangle) - ^1S_0
(|3\rangle)$,
if one prepares $^1D_2$ as the initial state.
Other candidates are metastable states of
noble gas atoms.

% ======================FIGURE 1 ====================================  
\begin{figure*}[htbp]
 \begin{center}
 \epsfxsize=0.5\textwidth
 \centerline{\epsfbox{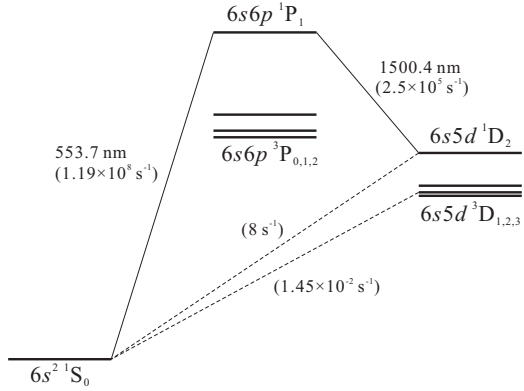}} \hspace*{\fill}
   \caption{Ba energy level. The transition time 69 s  of 
$^3D_2 \rightarrow ^1S_0$  (E2)   is taken from a theoretical
calculation of \cite{sr detection}.}
   \label{fig:energy_spectrum}
 \end{center} 
\end{figure*}
% ======================FIGURE====================================  

In the case of atoms coherently excited by a pulsed
laser the single photon superradiance
follows the stochastic spontaneous decay
\cite{dicke}, \cite{super-review}.
Hence its time profile of evolution is somewhat complicated.

Instead, we imagine in the present work that the initial excited state
is prepared by two laser irradiation, forming the dark state
\cite{dark state}, a pure mixture of two quantum
states $| 1\rangle $ and $| 3 \rangle$.
In the example of Ba the mixture of $^1D_2$ and $^1S_0$
is formed by two lasers of wavelengths,
554 nm and 1500 nm for Ba, corresponding to
transitions, $|1 \rangle \rightarrow |2 \rangle$ and
$|3 \rangle \rightarrow |2 \rangle$.
The time of the dark state formation
is of order, the larger of these E1 decay times, 
a few $\times 4 \mu$ sec,
which is shorter than two photon
superradiance time we consider below.
After the dark state formation we switch off
laser irradiation and measure two photon emission
until the collisional relaxation time 
($\sim $ 0.1 sec for gass targets).
This cycle is repeated as many times as possible.
We may thus expect for computations below that the dark 
state is present as the initial state
of our time development, and
may assume that a fraction of
the initial state is in the metastable
excited state $| 1 \rangle$ with a probability
$\sim$ (ratio of Rabi frequencies)$^2$.
This menas that for rate computation
at its maximum we may ignore
a complicated time profile, and
follow the S-matrix approach based on states on
the mass-shell \cite{density matrix}.

In more complicated situations in which two time
scales of dark state formation and superradiance
are comparable, one needs time integration of
the optical Bloch equation \cite{dark state}.

The emission rate, summed over target atom positions $\vec{r}$ of
two photon emission at $\vec{r}$ 
and detected at $\vec{r}_0$, is given by
(atoms distributed uniformly by a constant number density $n$)
\begin{eqnarray}
&&
\Gamma =
\int \left(
\Pi_i \frac{d^3 k_i}{(2\pi)^3}\right)
2\pi \delta(\Delta - \sum_i \omega_i)
\,| n \int_V d^3 r e^{i \sum_i \vec{k}_i\cdot(\vec{r} - \vec{r}_0)} 
{\cal M}( \vec{k}_i)|^2
\,,
\label{coherent rate 0}
\end{eqnarray}
where ${\cal M}( \vec{k}_i)$ is the probability amplitude
of emitting two photons of momenta $\vec{k}_i$
($\omega_i = |\vec{k}_i|$)
from a single atom.
Dependence on the detection point $\vec{r}_0$
disappears in the rate, and one may use a shape factor defined
by $F(\vec{K}) = \int_V d^3 r e^{i\vec{K}\cdot\vec{r}}$
(in general $\propto$ Fourier transform of the number density).

A cylindrical target is the standard example,
since preparation of excited atoms via laser irradiation
often gives a coherent region of this type.
The shape factor for a cylinder 
of area $\pi d^2$ and length $l$ \cite{eberly} is given by
\( \:
F(\vec{K})  = (4\pi \sin (K_z l/2)/K_z)
\left( de^{iK_{\rho}d}/(iK_{\rho}) - (1 - e^{iK_{\rho}d}) /K_{\rho}^2
\right)
\,.
\: \)
This function approaches the volume of the target,
$F(\vec{K}) \rightarrow \pi d^2 l$ in the long wavelength,
$|\vec{K}|^{-1} \ll d\,, l$.
For subsequent discussion it is convenient to
factor out volume related quantity and
introduce a dimensionless function ${\cal H}$ 
defined by ${\cal H} (dK_{\rho}\,, lK_z ) = |F(\vec{K})|^2/(\pi d^2 l)^2$,
which turns out
\begin{eqnarray}
&&
{\cal H}(x \,, y) = (\frac{4\sin (y/2)}{x^2 y})^2 
(2 - 2x \sin x - 2\cos x + x^2)
\,.
\end{eqnarray}
For small arguments of $x, y$,
${\cal H}(x \,, y) \sim 1 - x^2/18 - y^2/12$ to the second order.

We use the momentum vectors $\vec{k}_1 \,, \vec{k}_2$ 
of 2 photons which define a plane, not necessarily
containing the cylinder axis.
The total momentum component parallel to the cylinder axis
is denoted by $K_z$ and its orthogonal by $K_{\rho}$;
\begin{eqnarray}
&&
K_z = (\vec{k}_1 +  \vec{k}_2)\cdot\vec{e}_z
= \omega \cos \theta + (\Delta - \omega)\cos \theta_2
\,,
\\ &&
K_{\rho} = K \sin \theta_{12}
= \left( [(\vec{k}_1 +  \vec{k}_2)\cdot\vec{e}_x]^2 + 
[(\vec{k}_1 +  \vec{k}_2)\cdot\vec{e}_y]^2\right)^{1/2}
\nonumber \\ &&
\hspace*{-1cm}
= 
\left( \omega^2\sin^2 \theta  + (\Delta - \omega)^2\sin^2 \theta_2
+ 2 \omega(\Delta - \omega)\sin \theta \sin \theta_2 
\cos \varphi_{12} \right)^{1/2}
\,,
\end{eqnarray}
where the energy conservation $\Delta = \omega_1 + \omega_2$
is used.
The angles, $\theta\,,\theta_2\,, \varphi_{12}$, are
defined relative to the cylinder axis.

Two photon emission for the transiton
$^1D_2 \rightarrow ^1 S_0 $, may proceed via
intermediate state $| 2 \rangle = ^1 P_1$
($\Lambda-$type, hence $E_2 > E_1 > E_3$), 
and the double differential spectrum 
of a single photon is given by
\begin{eqnarray}
&&
\frac{d^2 \Gamma_{2\gamma}}{d\omega d \cos \theta}
= \int_0^{2\pi} d\varphi_{12} \int_{-1}^1 d\cos \theta_2
\frac{d^4 \Gamma_{2\gamma}}{d\omega d \cos \theta d \cos \theta_2 d \varphi_{12}}
\,,
\\ &&
\frac{d^4 \Gamma_{2\gamma}}{d\omega d \cos \theta d \cos \theta_2 d \varphi_{12}} =
\frac{(n d^2 l)^2 |d_{12}|^2 |d_{23}|^2 \omega^3 (\Delta - \omega)^3}{4 (2\pi)^2}
{\cal H}( dK_{\rho} \,, l K_z)
\nonumber \\ &&
\times
\left( \frac{1}{(\omega + \Delta_{21})^2 + \gamma^2/4} + 
\frac{1}{(\omega - \Delta_{23})^2 + \gamma^2/4}\right)
\,,
\label{2g dif-rate}
\end{eqnarray}
with $\Delta_{ij} = E_i - E_j$.
The dipole matrix element squared $|d_{ij}|^2$ 
for the transition $|i \rangle \leftrightarrow |j\rangle$ may be
replaced by the measurable E1 natural width $\gamma_{ij}$, using 
$\gamma_{ij} = |d_{ij}|^2 \Delta_{ij}^3/(3\pi)$.
The width factor in the formula (\ref{2g dif-rate}) $\gamma \approx
\gamma_{12} + \gamma_{23}$.

The rate becomes large  when the arguments $x\,,y$ of ${\cal H}(x\,, y)$
are of order unity or less.
This occurs only when two energy factors are close;
$\omega \approx \Delta - \omega$.
Hence a sharp peak appears at the half of the available
atomic energy; $\omega = \Delta/2$.

It is trivial to extend our result to the ladder-type
of atomic system such as Ba levels of
$^1D_2 \rightarrow ^3D_{2\,, 1\,,0} \rightarrow ^1S_0$.
What is needed is to replace a positive $\Delta_{21}$ by a negative value;
$\Delta_{21} = - \Delta_{12} < 0$.

\vspace{1cm}
{\bf Event rate and angular distribution}

% ======================FIGURE2====================================  
\begin{figure*}[htbp]
 \begin{center}
 %\begin{minipage}{6.5cm}
 \epsfxsize=\textwidth
 \centerline{\epsfbox{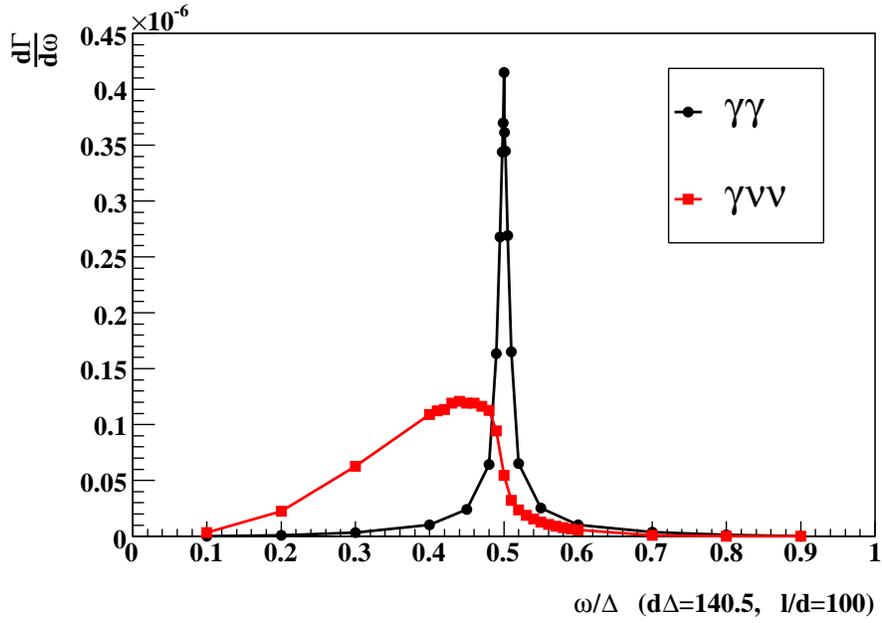}} \hspace*{\fill}
   \caption{Energy spectrum of MCTPE and radiative neutrino 
pair emission.
The rate of radiative neutrino pair emission
is rescaled up with a factor $1.0 \times 10^{41}$.
The size factors assumed are
$d  = 100 $eV$^{-1}$ and $l/d = 100$.}
   \label{fig:energy_spectrum}
 %\end{minipage}
 \end{center} 
\end{figure*}
% ======================FIGURE2====================================  

The integrated rate over the photon energy including the peak value
is of order, the central peak rate times the width factor around it,
which gives $O[\Delta \omega /(\pi dl\Delta^2)]$,
$1/(dl\Delta^2)$ arising from the Jacobian of variable change \\
$(\cos \theta_2 \,, \cos \varphi_{12}) 
\rightarrow (x\,, y)$.
With the expected $\Delta \omega = O[1/d]$, this factor becomes
$O[1/(V \Delta^2)]$, with $V$ the volume of target region.
Hence the integrated rate may be written as
$\Gamma_{2\gamma} = O[A n^2 V /\Delta^2]$.
More precisely,
a slight departure from the linearity $\propto V$  
is observed and is of logarithmic type; $V (1 + O[\ln d])$
The dimensionless quantity $A$ is intrinsic to the target atomic system,
and of order $1 \times 10^{-15}$ for the $\Lambda-$type Ba transition,
$^1D_2 \rightarrow ^1P_1 \rightarrow ^1S_0$.
Numerically, the rate at $n = 10^{10}cm^{-3}$ and $V = 1cm^3$ is 
$\sim 1.7 \times 10^{5}sec^{-1}$.
We show in Figure 2 the calculated energy spectrum along with one-photon
spectum arising from radiative neutrino pair emission
(see below).
The two photon process appears to fall into the range
of measurable region.

Dependence of the rate on target factors $\propto n^2 V$
differs from that for the single photon superradiance 
$\propto n^2 d^2 \lambda$ \cite{1g N-rate},
since the wavelength $\lambda$ limits the coherent region
in this case.
In the case of MCTPE there is no such limitation
of wavelength, because
the photon pair wave function 
$e^{i(\vec{k}_1 + \vec{k}_2)\cdot \vec{r}}$ can become of order unity
for the collinear momentum configuration $\vec{k}_1 + \vec{k}_2 = 0$.
The ratio of the single to the two photon superradiance is
$O[ 0.02 ](cm/l)$ for $^1 D_2$ of Ba atom \cite{1g 1-rate}.
Thus, one may be able to simultaneously
observe both of the single superradiance 
and MCTPE in a single experiment \cite{sr detection}.

It is interesting that for the ladder-type of system
there is a further enhancement if the resonance
condition, $\Delta/2 = \Delta_{23}$, namely 
$\Delta_{12} = \Delta_{23}$, is met.
This might open a new possibility of detecting
weak transition rates between magnetic levels
equally spaced energetically.
Ideal targets for the neutrino mass spectroscopy
might be atoms having the level structure of
ladder-type of $\Delta_{12} = \Delta_{23}$
such as vibrational levels of molecules.

% ======================FIGURE3====================================  
\begin{figure*}[htbp]
 \begin{center}
 %\begin{minipage}{6.5cm}
 \epsfxsize=\textwidth
 %\begin{minipage}{6.5cm}
 %\epsfxsize=\textwidth
 \centerline{\epsfbox{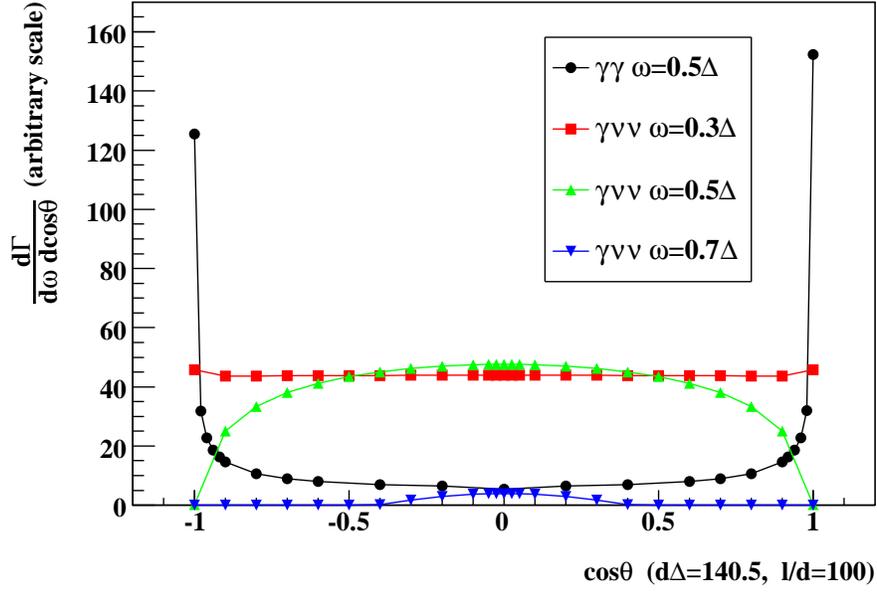}} \hspace*{\fill}
   \caption{
   Angular distribution of a photon from MCTPE at $\omega = \Delta/2$
and from radiative neutrino pair emission at $\omega = (0.3\,, 0.5\,, 0.7)
\Delta$, both measured
from the cylinder axis $\theta = 0$ and taking 
arbitrary rate units. 
The size factors assumed are $d  = 100 $eV$^{-1}$ and $l/d = 100$.}
   \label{fig:energy_spectrum}
 %\end{minipage}
 \end{center} 
\end{figure*}
% ======================FIGURE3==================================== 

The angular distribution of one photon
is symmetric under $\theta \leftrightarrow \pi - \theta$.
There are two angular components;
one isotropic and the other axial.
For the pencil-like cylinder $d \ll l$
the axial component is dominant and
we focuss on this component hereafter.
The axial component is limitted to a narrow angular region 
of $\theta \leq O[2d/l]$
in the forward and the backward direction,
as shown in Figure 3.
Both of them show the back to back correlation of two photons.

\vspace{1cm}
{\bf Radiative neutrino pair emission}

We now consider radiative neutrino pair emission;
$|1 \rangle \rightarrow |3 \rangle + \gamma
+ \nu_i \nu_j$.
The neutrino pair emission arises from the vertex of
four Fermi interaction of the kind,
$\nu_i^{\dagger} \nu_j^{\dagger} e^{\dagger}e$ and 
$\nu_i^{\dagger}\vec{\sigma} \nu_j^{\dagger} \cdot e^{\dagger}\vec{\sigma}e$, 
using two component spinor fields of $\nu_i$ and $e$.
Its precise form is determined by the standard electroweak
theory, and written out in \cite{my-06}.
The neutrino pair current involves both scalar 
$\nu_i^{\dagger} \nu_j^{\dagger}$ and
spin vector $\nu_i^{\dagger}\vec{\sigma}\nu_j^{\dagger}$ parts.
In the electron side the rate is largest
for a spin-flip magnetic type transiton of $e^{\dagger}\vec{\sigma}e$, 
when the orbital
wave function overlap is of order unity.

Thus, it is best in the Ba case to take the spin
triplet-singlet transition.
A convenient three level is  
$6s5d\, ^1D_2 - 6s5d\, ^3D_3 - 6s^2\, ^1S_0$ (ladder-type).
Other example includes 
fine structure
(FS) split levels of Yb;  $6s6p\, ^3P_2 - ^3P_1 - 6s^2\, ^1S_0$
and other alkali-earth atoms.
Since the possible background MCTPE contains a M1 transition,
the background rate is much smaller than two E1 case
in the preceding section.
In the following computations we ignore
the wave function overlap factor, which
is of order unity in magnetic type transitions.

A big difference from MCTPE is in the
way how momenta are balanced; the emitted photon is anti-parallel to
the neutrino pair. Since neutrinos are unobserved,
all neutrino momenta are integrated.
This gives a different size dependence, and
one expects from dimensional grounds a total maximal rate of order,
\( \:
G_F^2 \Delta^3 N^2/(\,(2\pi)^4 dl)
\sim 4  sec^{-1}(n/10^{20}cm^{-3})^2
d^3 l/cm^4
(\Delta/10 eV)^3
\,,
\: \)
where we took as an example $\Delta \geq O[10]eV$ relevant to noble gas atoms.
This estimate of rate is valid for the ladder-type of levels
for which the resonance enhancement is present, while
for the $\Lambda-$type there is a suppression
due to the off-resonance effect.

A straightforward computation similar to MCTPE
leads to a single photon differential
rate of the form,
\begin{eqnarray}
&&
%\hspace*{-1cm}
\frac{d^2 \Gamma_{\gamma \nu\nu}}{d\omega \, d\cos \theta}
=
\frac{4(n \pi d^2 l)^2G_F^2 |d_{23}|^2 \omega^3}
{(2\pi)^7\left((\omega - \Delta_{23})^2 + \gamma^2/4\right)} 
\int_{0}^{\sqrt{(\Delta - \omega)^2 - (m_i + m_j)^2}} dK\,
\times
\nonumber \\ &&
%\hspace*{2cm}
\int_0^{2\pi} d\varphi_K\,
\int_{-1}^1 d\cos \theta_K
{\cal H}(x\,, y) \frac{K^2}{(\Delta - \omega)^2 - K^2} \times
\nonumber \\ &&
\sqrt{\left((\Delta - \omega)^2 - K^2 - (m_i + m_j)^2\right)
\left( (\Delta - \omega)^2 - K^2 - (m_i - m_j)^2\right) } \times
\nonumber \\ &&
\hspace*{1cm}
\left(
k^{(0)}_{ij}f^{(0)}(\omega\,, K)
+ m_i^2 k^{(M)}_{ii}\delta_{ij}
\right)
\,,
\label{rnpe rate}
\\ &&
x = d
\left(
\omega^2 \sin^2 \theta + K^2 \sin^2 \theta_K 
+ 2K\omega \sin \theta \sin \theta_K \cos \varphi_K
\right)^{1/2} \,,
\\ &&
y = 
l\omega \cos \theta + lK \cos \theta_K \,,
\\ &&
f^{(0)}(\omega\,, K) = 
\left(
\frac{(\Delta - \omega)^2}{4} - \frac{K^2}{12} + \frac{1}{6}\frac{K^2}
{(\Delta - \omega)^2 - K^2}(m_i^2 + m_j^2)
\right.
\nonumber
\\ &&
\hspace*{1cm}
\left.
- \frac{1}{12}\frac{3(\Delta - \omega)^2 + K^2}
{\left( (\Delta - \omega)^2 - K^2 \right)^2}(m_i^2 - m_j^2)^2
\right)
\,,
\label{mj-dirac}
\\ &&
\hspace*{1cm}
k^{(0)}_{ij} =|c_{ij}^{(0)}|^2 + 3|c_{ij}^{(s)}|^2
\,, \hspace{0.5cm}
k^{(M)}_{ij} = |c_{ij}^{(0)}|^2 - 3|c_{ij}^{(s)}|^2 
\,,
\end{eqnarray}
where 
$k^{(0)}_{ij} =|c_{ij}^{(0)}|^2 + 3|c_{ij}^{(s)}|^2
\,, \hspace{0.2cm}
k^{(M)}_{ij} = |c_{ij}^{(0)}|^2 - 3|c_{ij}^{(s)}|^2 $ 
and $c^{(0)}_{ij} = U_{ei}^*U_{ej} \,, \hspace{0.2cm}
c^{(s)}_{ij} = U_{ei}^*U_{ej} - \frac{1}{2}\delta_{ij}$, 
and $U_{ei}$ is the neutrino mass mixing matrix element.
The second term in (\ref{rnpe rate}) $\propto k^{(M)}_{ii}$ 
represents the inteference term
proper to the identical Majorana particle.

In the radiative neutrino pair emission given by
(\ref{rnpe rate})
the dominant momentum region that gives small $x\,, y$
of ${\cal H}(x\,, y)$ is where the  
neutrino pair momentum is matched to that of the photon,
$\vec{K} \sim - \vec{k}_{\gamma}$.
This condition is readily obeyed,
if the pair momentum magnitude satisfies an inequality
$K \sim \omega \leq \sqrt{(\Delta - \omega)^2 - (m_i + m_j)^2}$,
the upper integration range of $K$.
This gives the threshold photon energy,
$\omega_{ij} = \Delta/2 - (m_i + m_j)^2/(2\Delta)$.
There are altogether six neutrino mass thresholds 
$\omega_{ij}\,, i\,,j = 1\,, 2\,, 3$.
A practical method of locating the neutrino mass
threshold is to observe the Jacobian peak of the threshold
at $\omega_{ij}$ as in the usual particle physics technique.
In Figure 2 the energy spectrum is shown for
values of $d \sim 20 \mu$m and $l \sim 2 $mm.
For this computation we assumed a single neutrino of mass 50 meV.
A successful discovery of the radiative neutrino pair
emission would immediately mean a highly sensitive
neutrino mass bound approaching $m_3 \sim 50$meV.

The total rate of radiative neutrino pair emission
also shows a weak logarithmic dependence on the cylinder size, 
$\propto V (1 + O[\ln (d)])$.
Angular distribution of radiative neutrino pair
emission is markedly different from MCTPE;
it has a broad structure, almost isotropic
for $\omega \leq \Delta/2$, around the orthogonal
direction to the cylinder axis, as shown in
Figure 3.
This may be used to distinguish the radiative
neutrino pair emission from MCTPE,
along with the broader energy spectrum.

\vspace{1cm}
{\bf Prospect for neutrino experiments}

Prospect for detecting radiative neutrino pair emission
is not bright, as it stands, for gaseous targets such as
those in a cell or atomic beam that can be excited
by laser irradiation.
To have a larger rate for measurement of
neutrino pair emission, it seems necessary
either to have a further enhancement mechanism or
to use denser targets such as atoms implanted within
a solid matrix or something similar.
The energy shift within solid does not cause
a problem, but it is necessary to have small width
broadening for successful neutrino mass spectroscopy.

For detection it is mandatory to repeat preparotory
laser irradiation in order to avoid the coherence loss
caused by collisional relaxation which destroys
metastable excited atoms and finally ends at the ground state.
Repeated cycles pumping from
the ground state into the excited state,
of repetition time, typically
$\Delta T = O$[100 ms] (relaxation time),
are needed.
A good energy resolution is also
crucial for performing the neutrino mass spectroscopy
to precisely locate the thresholds.

Further study on these issues is obviously
required prior to actual experiments.


\begin{thebibliography}{99}
\bibitem{density matrix}
Alternatively, one may compute the minus of the survival rate
of a macrocoscopic N atomic state 
$| I \rangle$, $- \frac{d}{dt} \langle I | \rho | I \rangle$
against two photon emission, 
using the reduced density matrix $\rho$
integrated over envirornment variables of
radiation and neutrino fields out of the entire dynamical
variables of atoms + fields.
This computation, to the leading order of perturbation theory,
yields the real part of time integration of the product,
two photon propagators times transition matrix element squared,
$\int_{-\infty}^{t}dt' D_1 (\vec{r}_a - \vec{r}_b\,, t-t')
D_2 (\vec{r}_a - \vec{r}_b\,, t-t') e^{-i \Delta(t-t')}
|{\cal M}(\vec{k}_i)|^2$
where $D_i (\vec{r}_a - \vec{r}_b\,, t-t')$ 
is the photon propagator of momentum $\vec{k}_i$,
namely $i e^{i\vec{k}_i\cdot(\vec{r}_a - \vec{r}_b) - i \omega_i (t-t')}
\theta(t-t') /(2\omega_i)$, 
the photon being emitted at atomic location $\vec{r}_b$ and absorbed
at $\vec{r}_a$.
The energy conservation factor in eq.(\ref{coherent rate 0}) 
$ 2\pi \delta(\Delta - \sum_i \omega_i)$ 
and  the shape factor in the text are thus derived.
This derivation is explained in standard reviews such as \cite{super-review}.
\bibitem{1g N-rate}
The total rate, calculated using our ${\cal H}(x\,, y)$,
for the superradiant single photon emission is given by
$(54 d^2 n^2 \gamma)\ln (2 d^2 \Delta^2/9)/(\pi \Delta^4)$ 
with $\gamma$ the E1 or M1 rate $\propto \Delta^3$ from a single atom.
\end{thebibliography}
\end{document}